\documentclass[utf8]{FrontiersinHarvard}
\usepackage{url,microtype,subcaption}
\usepackage[onehalfspacing]{setspace}
\usepackage[varg]{txfonts}
\usepackage{amsmath}
\usepackage{graphicx}
\usepackage{comment}
\usepackage{color} 
\usepackage{natbib}
\usepackage{threeparttable}

\newcommand{\fig}[1]{Fig.~\ref{#1}}
\newcommand{\tbl}[1]{Table~\ref{#1}}
\newcommand{\sect}[1]{Sect.~\ref{#1}}
\newcommand{\equ}[1]{Equation~(\ref{#1})}

\newcommand{\apj}{Astrophysical Journal.}
\newcommand{\apjl}{Astrophysical Journal Letters.}
\newcommand{\solphys}{Solar Physics.}
\newcommand{\aap}{Astronomy and Astrophysics.}
\newcommand{\jgr}{Journal of Geophysics Research.}
\newcommand{\ssr}{Space Science Reviews.}
\newcommand{\nar}{New Astronomy Review.}

\def\keyFont{\fontsize{8}{11}\helveticabold }
\def\firstAuthorLast{Quanhao Zhang {et~al.}} 
\def\Authors{Quanhao Zhang\,$^{1,2,3,*}$, Xin Cheng\,$^{4}$, Rui Liu\,$^{2,3,5}$, Anchuan Song\,$^{2,3,6}$, Xiaolei Li\,$^{2,3,6}$, and Yuming Wang\,$^{1,2,3}$}
\def\Address{$^{1}$Deep Space Exploration Laboratory/School of Earth and Space Sciences, University of Science and Technology of China, Hefei 230026, China \\
$^{2}$CAS Key Laboratory of Geospace Environment, School of Earth and Space Sciences, University of Science and Technology of China, Hefei 230026, China\\
$^{3}$CAS Center for Excellence in Comparative Planetology, University of Science and Technology of China, Hefei 230026, China\\ 
$^{4}$School of Astronomy and Space Science, Nanjing University, Nanjing 210023, China\\ 
$^{5}$Collaborative Innovation Center of Astronautical Science and Technology, Hefei, Anhui 230026, China\\ 
$^{6}$Mengcheng National Geophysical Observatory, School of Earth and Space Sciences, University of Science and Technology of China, Hefei 230026, China }
\def\corrAuthor{Quanhao Zhang}
\def\corrEmail{zhangqh@ustc.edu.cn}

\begin{document}
\onecolumn
\firstpage{1}

\title[Influence of reconnection on catastrophe]{Influence of magnetic reconnection on the eruptive catastrophes of coronal magnetic flux ropes} 

\author[\firstAuthorLast ]{\Authors} 
\address{} 
\correspondance{} 

\extraAuth{}

\maketitle

\begin{abstract}
\section{}
Large-scale solar eruptive activities have a close relationship with coronal magnetic flux ropes. Previous numerical studies have found that the equilibrium of a coronal flux rope system could be disrupted if the axial magnetic flux of the rope exceeds a critical value, so that the catastrophe occurs, initiating the flux rope to erupt. Further studies discovered that the catastrophe does not necessarily exist: the flux rope system with certain photospheric flux distributions could be non-catastrophic. It is noteworthy that most previous numerical studies are under the ideal magnetohydrodynamic (MHD) condition, so that it is still elusive whether there is the catastrophe associated with the critical axial flux if magnetic reconnection is included in the flux rope system. In this paper, we carried out numerical simulations to investigate the evolutions of coronal magnetic rope systems under the ideal MHD and the resistive condition. Under the ideal MHD condition, our simulation results demonstrate that the flux rope systems with either too compact or too weak photospheric magnetic source regions are non-catastrophic versus varying axial flux of the rope, and thus no eruption could be initiated; if there is magnetic reconnection in the rope system, however, those flux rope systems could change to be capable of erupting via the catastrophe associated with increasing axial flux. Therefore, magnetic reconnection could significantly influence the catastrophic behaviors of flux rope system. It should be both the magnetic topology and the local physical parameters related to magnetic reconnection that determine whether the increasing axial flux is able to cause flux rope eruptions. 
\tiny
\keyFont{ \section{Keywords:} Sun: filaments -- Sun: prominences -- Sun: flares -- Sun: coronal mass ejections -- Sun: magnetic fields -- Sun: activity} 
\end{abstract}

\section{Introduction}

Large-scale solar eruptive activities are the principal cause of extreme space weather in Earth and planetary space environments \citep{svestka2001a,Cheng2014,Lugaz2017,Gopalswamy2018a}. Different kinds of large-scale solar eruptions, including prominence/filament eruptions \citep{Li2016,Jenkins2018,Fan2020,Devi2021}, flares \citep{Chen2015,Li2016a,Cheung2019}, and coronal mass ejections \cite[CMEs,][]{Shen2014,Lamy2019,Bemporad2022}, are generally considered to be different manifestations of the eruptions of coronal magnetic flux rope \citep{Zhang2001,Vrvsnak2005a,Jiang2016a,Chen2020a,Liu2020a}. Therefore, investigating the initiation and evolution of flux rope eruptions is crucial not only for understanding solar eruptions, but also for space weather forecasting. Many theoretical models have been proposed to shed light on the physical scenario of coronal flux rope eruptions. These models are based on different kinds of physical mechanisms, such as ideal magnetohydrodynamic(MHD) instabilities \citep{Torok2003a,Aulanier2010a,Guo2010,Savcheva2012b,Keppens2019}, magnetic reconnection \citep{Antiochos1999a,Chen2000a,Moore2001a,Sterling2004,Archontis2008b,Inoue2015}, and the catastrophes of coronal flux ropes \citep{vanTend1978a,Forbes1995a,Lin2002b,Kliem2014,Zhang2021}.
\par
Many theoretical studies suggested that flux rope catastrophes are intriguing candidates for the source of solar eruptions \citep[e.g.,][]{Lin2000a,Isenberg1993a}. In the flux rope catastrophe theory, the onset of the eruption is approached as the loss of equilibrium of the coronal flux rope system. Before the onset, the flux rope should be static or quasi-static \citep{Toeroek2013,Liu2020a}, indicating that the flux rope system is in equilibrium. If the equilibrium is not disrupted, the net force on the flux rope will always be zero, so that its state of motion will remains unchanged, i.e., no eruption of the flux rope could occur. Therefore, loss of equilibrium must occur during the onset of the eruption, and the resultant net force initiates the flux rope to erupt. This is the fundamental scenario of the catastrophe of coronal flux ropes. The investigation about when and how the catastrophe occurs is based on the equilibrium manifold \citep{Forbes1995a,Isenberg2007a,Kliem2014}, which consists of all the equilibrium states of the flux rope system. For example, by analytically deriving the equilibrium manifold as a function of current within the flux rope, \cite{vanTend1978a} for the first time discovered that there is a critical value of the current, beyond which there are no neighbouring equilibrium states so that the catastrophe occurs, which results in a discontinuous equilibrium manifold; this critical value is referred to as the catastrophic point. Many more analytical studies have verified that the catastrophe could occur in various types of coronal flux rope systems, resulting in the eruption of the flux rope \citep[][]{Priest1990a,Forbes1995a,Lin2000a,Demoulin2010a,Longcope2014a,Kliem2014}. In addition, numerical simulations have also been carried out in many previous studies to investigate the catastrophes of coronal flux ropes \citep[][]{Forbes1990a,Chen2007a,Zhang2017a,Zhang2021}. In numerical studies, the equilibrium manifold as a function of a certain physical parameter is obtained by simulating the equilibrium solutions of the flux rope system with different values of this parameter. For example, \cite{Su2011a} discovered that the equilibrium manifold versus the axial magnetic flux of the flux rope is discontinuous at a critical value, beyond which a catastrophe occurs. Similar results are obtained in many other numerical studies \citep[][]{Bobra2008,Su2009,Su2011a,Zhang2017a,Zhang2017}, indicating that the axial flux of the rope should play an important role in initiating solar eruptions.
\par
The catastrophe associated with the critical axial flux does not necessarily exist in coronal flux rope system. Previous numerical studies have demonstrated that the photospheric flux distributions of the background field greatly influence the catastrophic behaviors of coronal flux rope systems \citep{Sun2007a,Zhang2017a,Zhang2017}. Previous studies found that there are two types of non-catastrophic flux rope systems: if the photospheric flux is too concentrated toward the polarity inversion line (hereafter ``compact'' cases), or the total magnetic flux originating from the photospheric magnetic source regions is too weak (hereafter ``weak'' cases), the equilibrium manifold as a function of the rope's axial flux will be continuous so that no catastrophe could occur. It is still an open question why there is no catastrophe in these two types of flux rope system; \cite{Zhang2017} inferred that the constraint from the background field on the flux rope might probably plays an considerable role. It is noteworthy that magnetic reconnection is completely prohibited in those studies, so that their conclusion about the two types of non-catastrophic flux rope systems is under the ideal MHD condition. Since the actual solar corona is resistive, there should be reconnection in actual coronal flux rope systems \citep{Jiang_2021,Yan2022,Bian2022}. In fact, magnetic reconnection could not only change the magnetic topology, resulting in the redistribution of the Amp\`{e}re's force (also known as Lorentz force in many papers), but the reconnection outflow could also push the flux rope upward \citep{Chen2004,Wang2007,Xue2016,Cheng2020a}. This indicates that magnetic reconnection could contribute to the force balance of a coronal flux rope system, implying that the catastrophic behaviors of the rope system might differ under the ideal MHD and the resistive condition. Therefore, it is critical to assess the influence of magnetic reconnection on the catastrophic behaviors of coronal flux rope systems, so as to shed more light on the catastrophe theory for solar eruptions. To achieve this, we use a 2.5-dimensional MHD numerical model to simulate the catastrophic behaviors of the coronal flux rope systems with either compact or weak photospheric source regions. Both the equilibrium manifolds versus the axial magnetic flux under the ideal MHD and the resistive condition are simulated, based on which the influence of magnetic reconnection is investigated. The rest of this paper is arranged as follows: the numerical model and simulating procedures are introduced in \sect{sec:model}, the simulation results are presented in \sect{sec:result}, and the conclusion and discussion are given in \sect{sec:dc}.

\section{Numerical model}
\label{sec:model}

\subsection{Basic equations}
\label{sec:equations}
The numerical model in our simulation is similar as those used in \cite{Zhang2017a,Zhang2017}. For 2.5-dimensional cases, by assuming all the quantities satisfy $\partial/\partial z=0$, the magnetic field can be written in form of magnetic flux function $\psi$:
\begin{equation}
\textbf{B}=\triangledown\psi\times\hat{\textbf{\emph{z}}}+B_z\hat{\textbf{\emph{z}}}.\label{equ:mf}
\end{equation}
Here $\textbf{\emph{z}}$ is the unit vector in $z-$direction. Obviously, the divergence-free condition ($\triangledown\cdot \textbf{B}=0$) is always satisfied. With the form given in \equ{equ:mf}, the MHD equations in our simulations could be rewritten in the following form:\par
\begin{align}
&\frac{\partial\rho}{\partial t}+\triangledown\cdot(\rho\textbf{\emph{v}})=0,\label{equ:cal-st}\\
\nonumber &\frac{\partial\textbf{\emph{v}}}{\partial t}+\frac{2}{\rho\beta_0}(\vartriangle\psi\triangledown\psi+B_z\triangledown B_z+\triangledown\psi\times\triangledown B_z)+\textbf{\emph{v}}\cdot\triangledown\textbf{\emph{v}}\\ 
&~~~+\triangledown T +\frac{T}{\rho}\triangledown\rho+g\hat{\textbf{\emph{y}}}=0,\\
&\frac{\partial\psi}{\partial t}+\textbf{\emph{v}}\cdot\triangledown\psi-\eta\vartriangle\psi=0,\\
&\frac{\partial B_z}{\partial t}+\triangledown\cdot(B_z\textbf{\emph{v}})+(\triangledown\psi\times\triangledown v_z)\cdot\hat{\textbf{\emph{z}}}-\eta\vartriangle B_z=0,\\
\nonumber &\frac{\partial T}{\partial t}-\frac{\eta(\gamma-1)}{\rho R}\left[(\vartriangle\psi)^2+|\triangledown\times(B_z\hat{\textbf{\emph{z}}})|^2 \right]\\
&~~~+\textbf{\emph{v}}\cdot\triangledown T +(\gamma-1)T\triangledown\cdot\textbf{\emph{v}}=0,\label{equ:cal-en}
\end{align}
where
\begin{equation}
\vartriangle\psi=\frac{\partial^2\psi}{\partial x^2}+\frac{\partial^2\psi}{\partial y^2},~~\vartriangle B_z=\frac{\partial^2 B_z}{\partial x^2}+\frac{\partial^2 B_z}{\partial y^2}.
\end{equation}
Here $\rho$ is the density, $\textbf{\emph{v}}$ the velocity, $T$ the temperature, $\gamma=5/3$ the polytropic index, $g$ the gravity, $\eta$ the resistivity, and $\beta_0=0.1$ the characteristic ratio of the gas pressure to the magnetic pressure, which is comparable to the value in quiescent regions \citep{Anzer2007,Hillier2012,Xia2012}. The $x$, $y$, and $z-$component of the quantities are denoted by the subscript $x, y, z$. The radiation and the heat conduction in the energy equation are neglected. The numerical domain in our simulation is $0<x<200$ Mm, $0<y<500$ Mm; it is discretized into 200$\times$250 uniform meshes.

\subsection{Initial state}
\label{sec:initial}
\begin{figure*}
\includegraphics[width=\hsize]{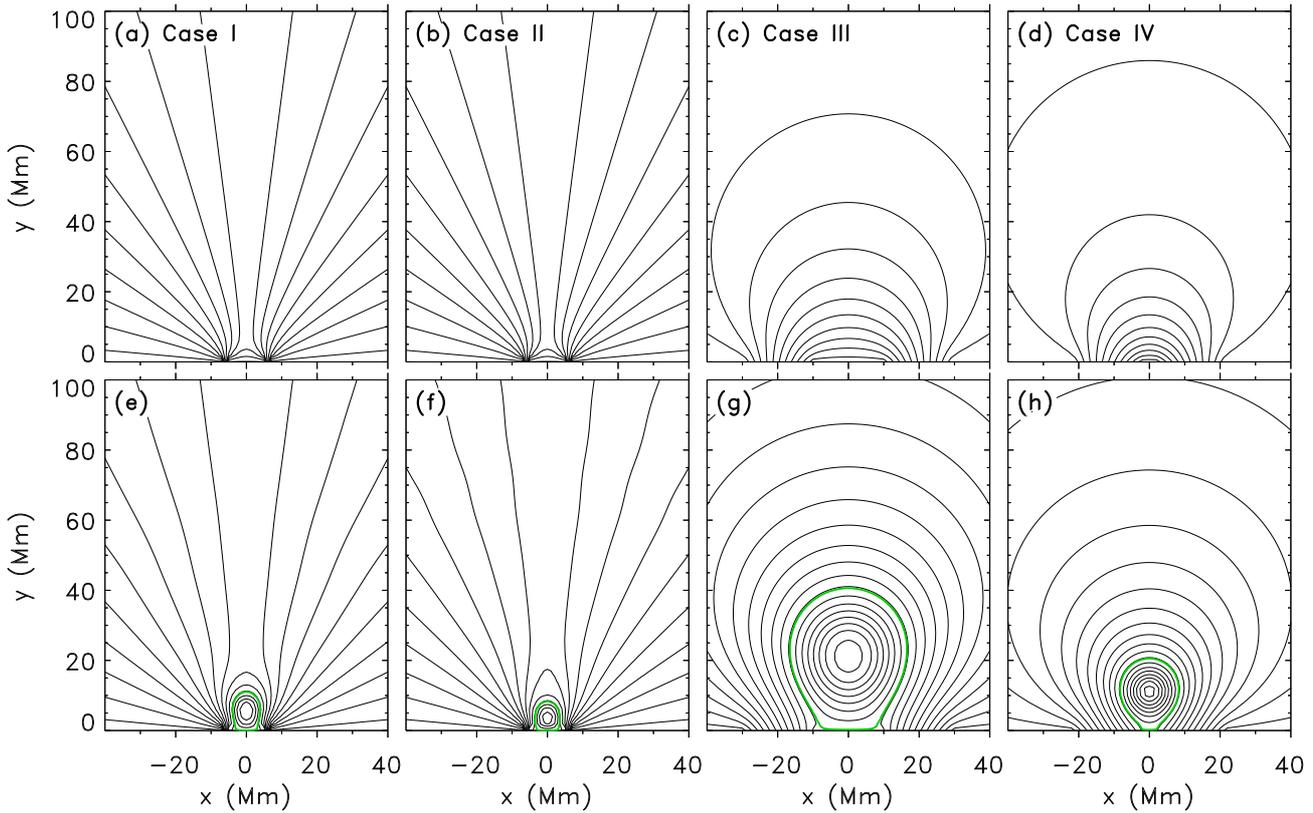}
\caption{The magnetic configurations of the background fields and initial states in the simulation. Panel (a) is the background field for Case I, and panel (e) is the corresponding initial state; panels (b) and (f), (c) and (g), (d) and (h) are those for Case II, III, IV, respectively. The black curves are the magnetic field lines, and the green curves mark the boundary of the flux rope. The physical parameters for the background fields and initial states are tabulated in \tbl{tbl:para}.}\label{fig:ini}
\end{figure*}
The initial states in our simulations are constructed by numerical procedures. Here we first use the complex variable method \citep[e.g.,][]{Hu2000a,Zhang2017a} to construct the background field originating from photospheric magnetic source regions. The background field is a partially open bipolar field, with a negative surface magnetic charge located at the lower boundary within $-b<x<-a$, and a positive one within $a<x<b$; the distance between the two source regions is $D=2a$, and the width of the source region $W=b-a$. The background field in $x-y$ plane could then be expressed in the complex variable form as:
\begin{equation}
f(\omega)\equiv B_x-iB_y=\lambda\frac{(\omega+iy_N)^{1/2}(\omega-iy_N)^{1/2}}{F(a,b,y_N)}\mathrm{ln}\left( \frac{\omega^2-a^2}{\omega^2-b^2}\right),\label{equ:complex}
\end{equation}
where $\omega=x+iy$,\par
\begin{align}
\nonumber &F(a,b,y_N)=\frac{1}{b-a}\int_a^b(x^2+y_N^2)^{1/2}dx=\frac{1}{2(b-a)}\times\\ &\left[b(b^2+y_N^2)^{1/2}-a(a^2+y_N^2)^{1/2}+y_N^2\mathrm{ln}\left(\frac{b+(b^2+y_N^2)^{1/2}}{a+(a^2+y_N^2)^{1/2}} \right)\right],
\end{align}
and the neutral point of the background field is at ($y=y_N$, $x=0$). As indicated by \equ{equ:complex}, the magnetic field strength of the background field is proportional to the dimensionless parameter $\lambda$. The magnetic flux function within the domain could then be calculated by:
\begin{equation}
\psi(x,y)=\mathrm{Im}\left\lbrace\int f(\omega)d\omega \right\rbrace,\label{equ:integral}
\end{equation}
With calculated the flux function $\psi$, and by letting $B_z=0$ in the background field, the magnetic configuration of the background field is obtained (\equ{equ:mf}). In particular, the flux function of the background field at the lower base, $\psi_i$, could be calculated as:
\begin{equation}
 \psi_i=\psi(x,0) = \left\{
              \begin{array}{ll}
              {\lambda\widehat{\psi_0}\pi(b-a)}, &{|x|<a}\\
              {\lambda\widehat{\psi_0}\pi(b-a) F(|x|,b,y_N)/F(a,b,y_N)}, &{a\leqslant|x|\leqslant b}\\
              {0}, &{|x|>b}
              \end{array}  
         \right.\label{equ:fluxb}
\end{equation}
and the flux function at the neutral point of the background field is
\begin{equation}
 \psi_N=\psi(0,y_N)=\frac{\lambda\widehat{\psi_0}\pi(b^2-a^2)}{2F(a,b,y_N)}
\end{equation}
Here $\psi_c=\lambda\pi(b-a)\widehat{\psi_0}$ ($\widehat{\psi_0}=37.25\mathrm{~Mx~cm^{-2}}$) is the total magnetic flux emanating upward from the positive photospheric source region per unit length along $z-$direction, i.e. $\psi_c$ represents the intensity of background magnetic field originating from the photospheric source regions. The ratio of the background field's open magnetic flux to its total magnetic flux could then be calculated by:
\begin{equation}
\kappa=\frac{\psi_N}{\psi_c}.
\end{equation}
For a given background magnetic field, its characteristic physical parameters include $D$, $W$, $\psi_c$ and $\kappa$, which could be calculated with a group of given $a$, $b$, $y_N$, and $\lambda$. For example, by letting $a=5.00$ Mm, $b=7.00$ Mm, $y_N=4.49$ Mm, $\lambda=1.0$, the calculated background configuration is plotted in \fig{fig:ini}(a), and the characteristic parameters of this background field are: $D=10$ Mm, $W=2$ Mm, $\psi_c=2.34\times10^{10}$ Mx cm$^{-1}$, and $\kappa=0.80$, as tabulated in the second column (Case I) in \tbl{tbl:para}. The initial corona in our simulations is static and isothermal:
\begin{equation}
T_c\equiv T(0,x,y)=1\times10^6 ~\mathrm{K},\ \  \rho_c\equiv\rho(0,x,y)=\rho_0\mathrm{e}^{-gy},\label{equ:rhot}
\end{equation}
where $\rho_0=3.34\times10^{-13}$ kg~m$^{-3}$.
\par
With the background field obtained above, we let a flux rope emerge from the lower base of the domain with similar simulating procedures as those introduced in, e.g., \cite{Zhang2017a} and \cite{Zhang2020}, and then relax the flux rope system to equilibrium. The resultant equilibrium state consisting of a flux rope embedded in the background field is the constructed initial state. For the background field plotted in \fig{fig:ini}(a), the constructed initial state is illustrated in \fig{fig:ini}(e). The magnetic properties of a flux rope could be characterized by its magnetic fluxes, including the total axial flux passing through the rope's cross section, $\Phi_z=\iint B_z\mathrm{d}S$, and the annular flux per unit length along $z$-direction of the rope, $\Phi_p$; in 2.5 dimensional cases, the poloidal flux is calculated as:
\begin{equation}
\Phi_p=\psi_{rc}-\psi_{rb},\label{equ:phip}
\end{equation}
where $\psi_{rc}$ and $\psi_{rb}$ are the flux function $\psi$ at the center and the boundary of the flux rope, respectively. For the initial state illustrated in \fig{fig:ini}(e), $\Phi_{z0}=1.416\times10^{19}$ Mx and $\Phi_{p0}=7.450\times10^{9}$ Mx cm$^{-1}$. With this initial state, we could further simulate the catastrophic behaviors of the coronal flux rope system with the given photospheric magnetic conditions (detailed simulating procedures are introduced in \sect{sec:method}). It is noteworthy that the radius of the flux rope in our simulation is finite, so that the thin-rope approximation is not satisfied. Under this circumstance, the initial state could only be obtained by numerical procedures.
\par
In our simulations, apart from those plotted in \fig{fig:ini}(a) and \fig{fig:ini}(e), we also construct another three groups of background field and initial state with similar procedures introduced above, and their parameters are tabulated in \tbl{tbl:para}; all these 4 cases are marked as Case I$\sim$IV, respectively. The background field and the initial state in Case II are illustrated in \fig{fig:ini}(b) and \fig{fig:ini}(f), respectively. As shown by the the second and the third column in \tbl{tbl:para}, $D$, $W$, and $\kappa$ in Case I and Case II are the same; the only difference in the characteristic parameters of the background field between these two cases is that $\psi_c$ in Case II is much smaller than that in Case I (achieved by adjusting $\lambda$ to 0.1 in Case II), i.e. the intensity of the background magnetic field originating from the photospheric source regions in Case II is much weaker than those in Case I. For Case III and Case IV, the corresponding background fields and initial states are illustrated in \fig{fig:ini}(c)-\ref{fig:ini}(d) and \fig{fig:ini}(g)-\ref{fig:ini}(h), respectively. As shown in the last two columns in \tbl{tbl:para}, the characteristic parameters of the background field in these two cases are the same, except that the distance between the source regions, $D$, is much smaller in Case IV than that in Case III. 
\begin{table}
\caption{The parameters of the background fields and the initial states for Case I-IV.}
\label{tbl:para}
\centering
\begin{threeparttable}
\renewcommand{\arraystretch}{1.5}
\begin{tabular}{|l|c|c|c|c|}
\hline
 & Case I & Case II & Case III & Case IV \\
\hline
$a$ (Mm)    & 5.00 & 5.00 & 10.00   & 2.00 \\
$b$ (Mm)    & 7.00 & 7.00 & 30.00   & 22.00\\
$y_n$ (Mm)  & 4.49 & 4.49 & 1999.89 & 1199.93\\
$\lambda$   & 1.00 & 0.10 & 0.10    & 0.10 \\\hline
$D$ (Mm)    & 10.00& 10.00& 20.00   & 4.00\\
$W$ (Mm)    & 2.00 & 2.00 & 20.00   & 20.00\\
$\psi_c$ (Mx cm$^{-1}$)& 2.340$\times10^{10}$ & 0.234$\times10^{10}$ & 2.340$\times10^{10}$ & 2.340$\times10^{10}$\\
$\kappa$    & 0.80 & 0.80 & 0.01    & 0.01\\\hline
$\Phi_{z0}$ (Mx) & 1.416$\times10^{19}$ & 1.118$\times10^{18}$ & 5.960$\times10^{19}$ & 2.421$\times10^{19}$ \\
$\Phi_{p0}$ (Mx cm$^{-1}$) & $7.450\times10^{9}$ & $7.450\times10^{8}$ & $1.490\times10^{10}$ & $1.490\times10^{10}$ \\
\hline
 \end{tabular} 
  \begin{tablenotes}
   \footnotesize
        \item[*] Here $a$, $b$, $y_n$, and $\lambda$ are the parameters used to construct the background field; $D$, $W$, $\psi_c$, and $\kappa$ are the corresponding characteristic physical parameters of the background field; $\Phi_{z0}$ and $\Phi_{p0}$ are the magnetic fluxes of the initial flux rope.
  \end{tablenotes}
 \end{threeparttable}
\end{table}

\subsection{Simulating procedures}
\label{sec:method}
Starting from each of the four initial states obtained above, we simulate the equilibrium states of the corresponding flux rope system with different axial magnetic fluxes of the rope: from $0\sim20\tau_A$ ($\tau_A=17.4$ s is the typical Alfv\'{e}n transit time), the axial flux $\Phi_z$ and the poloidal flux $\Phi_p$ are adjusted from the initial values to a certain group of target values ($\Phi_z^1$, $\Phi_p^t$), and from $20\sim300\tau_A$, the magnetic system is relaxed to equilibrium, with the magnetic fluxes fixed at the target values ($\Phi_z^1$, $\Phi_p^t$). We note that the conservation of the poloidal flux is achieved by fixing the flux function at the rope center, $\psi_{rc}$, during the relaxation (\equ{equ:phip}). The final state at $t=300\tau_A$ is regarded as the equilibrium state of the flux rope system for $\Phi_z^1$. During the relaxation, magnetic reconnection is either included or completely prohibited during the relaxation, corresponding to the resistive condition and ideal MHD condition, respectively. For the simulations under the resistive condition, anomalous resistivity is used:\par
\begin{align}
\eta=
\begin{cases}
0,& ~j\leq j_c\\
\eta_m\mu_0v_0L_0(\frac{j}{j_c}-1)^2.& ~j> j_c \label{equ:res}
\end{cases}
\end{align}
Here $\eta_m$=0.10, $L_0=10^7$ m, $v_0=128.57$ km s$^{-1}$, $j_c=2.96\times10^{-5}$ A m$^{-2}$, and $\mu_0$ is the vacuum magnetic permeability. On the other hand, for the simulations under the ideal MHD conditions, we use similar simulating procedures as those in \cite{Zhang2017a,Zhang2017} during the relaxation to prohibit the reconnection: first set the resistivity $\eta$ to be $0$, and then reassign the flux function $\psi$ along the current sheet with the initial values at each time step, so as to keep $\psi$ invariant along the current sheet. Since any reconnection will reduce the value of $\psi$ at the current sheet, both physical and numerical reconnections are prohibited with the simulating procedures introduced above. For the given target values $\Phi_z^1$, we are able to simulate two equilibrium states with the simulating procedures above: if reconnection is prohibited during the relaxation, the resultant equilibrium state is under ideal MHD condition, whereas the resultant equilibrium state is under resistive condition if reconnection is included during the relaxation. 
\par
Similar simulations are repeated for different target values (($\Phi_z^2$, $\Phi_p^t$), ($\Phi_z^3$, $\Phi_p^t$), ...) so as to obtain equilibrium states with different $\Phi_z$. Eventually, all the calculated equilibrium states without reconnection constitute a equilibrium manifold under the ideal MHD condition, and all those with reconnection constitute the equilibrium manifold under the resistive condition. 
\par
In our simulations, the quantities at the lower boundary of the domain are fixed, so that the lower boundary corresponds to the photosphere. Symmetric boundary condition is used at the left side of the domain ($x=0$). At the other boundaries, increment equivalent extrapolation \citep[][]{Zhang2017a,Zhang2017} is used to prescribe the boundary quantities:
\begin{equation}
U^{n+1}_{b}=U^{n+1}_{b-1}+U^{n}_{b}-U^{n}_{b-1},
\end{equation}
where $U$ represents the quantities. The superscript $n$ denote the quantities at the current time step, and $n+1$ the next time step; the subscript $b$ indicates the quantities at the boundary, and $b-1$ those at the grids next to the boundary.

\section{Simulation results}
\label{sec:result}

\subsection{Weak cases}
\label{sec:weak}
\begin{figure*}
\includegraphics[width=\hsize]{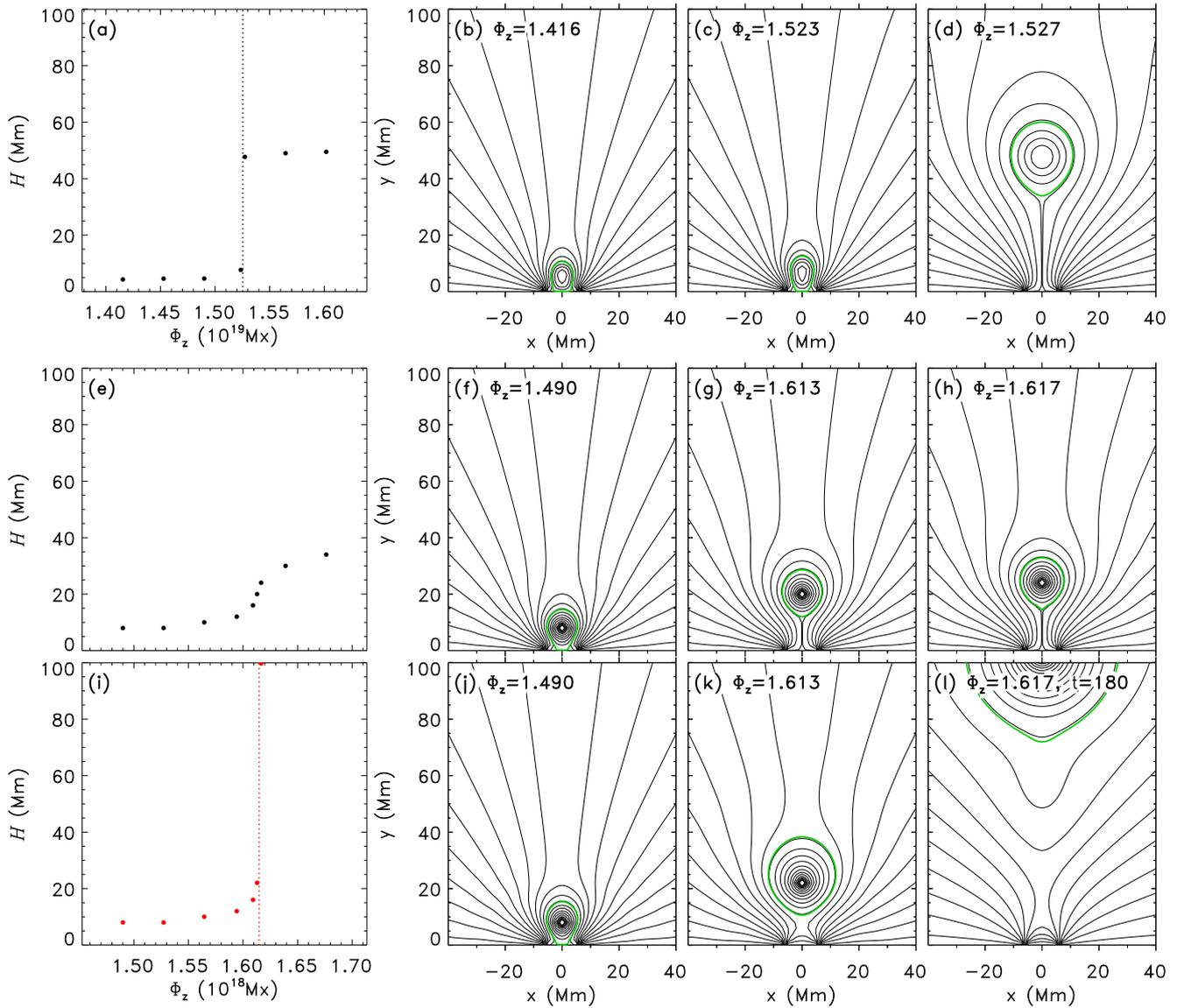}
\caption{The simulation results for Case I and Case II. Panel (a) is the equilibrium manifold for Case I under the ideal MHD condition; panels (b)-(d) are the magnetic configuration of the simulated equilibrium states for Case I, and the corresponding $\Phi_z$ is marked at the top of each panel. Panels (e)-(h) and panels (i)-(l) are the simulation results for Case II under the ideal MHD and the resistive condition, respectively; the red vertical dotted line in panel (i) denote the catastrophic point $\Phi_{zc}^w$. The green curves mark the boundary of the flux rope.}\label{fig:weak}
\end{figure*}

The equilibrium manifold as a function of the axial flux $\Phi_z$ under the ideal MHD condition for Case I is plotted in \fig{fig:weak}(a). All the equilibrium states have the same $\Phi_p=7.45\times10^9$ Mx cm$^{-1}$. Obviously, this equilibrium manifold is discontinuous: the flux rope keeps sticking to the photosphere (\fig{fig:weak}(b)-\ref{fig:weak}(c)) before reaching a critical axial flux at about $1.527\times10^{18}$ Mx, which is marked by the black vertical dotted line in \fig{fig:weak}(a); otherwise, the flux rope jumps upward (\fig{fig:weak}(d)), indicating that a catastrophe occurs. Therefore, the flux rope system in Case I is catastrophic under the ideal MHD condition.
\par
\begin{figure*}
\centering
\includegraphics[width=0.8\hsize]{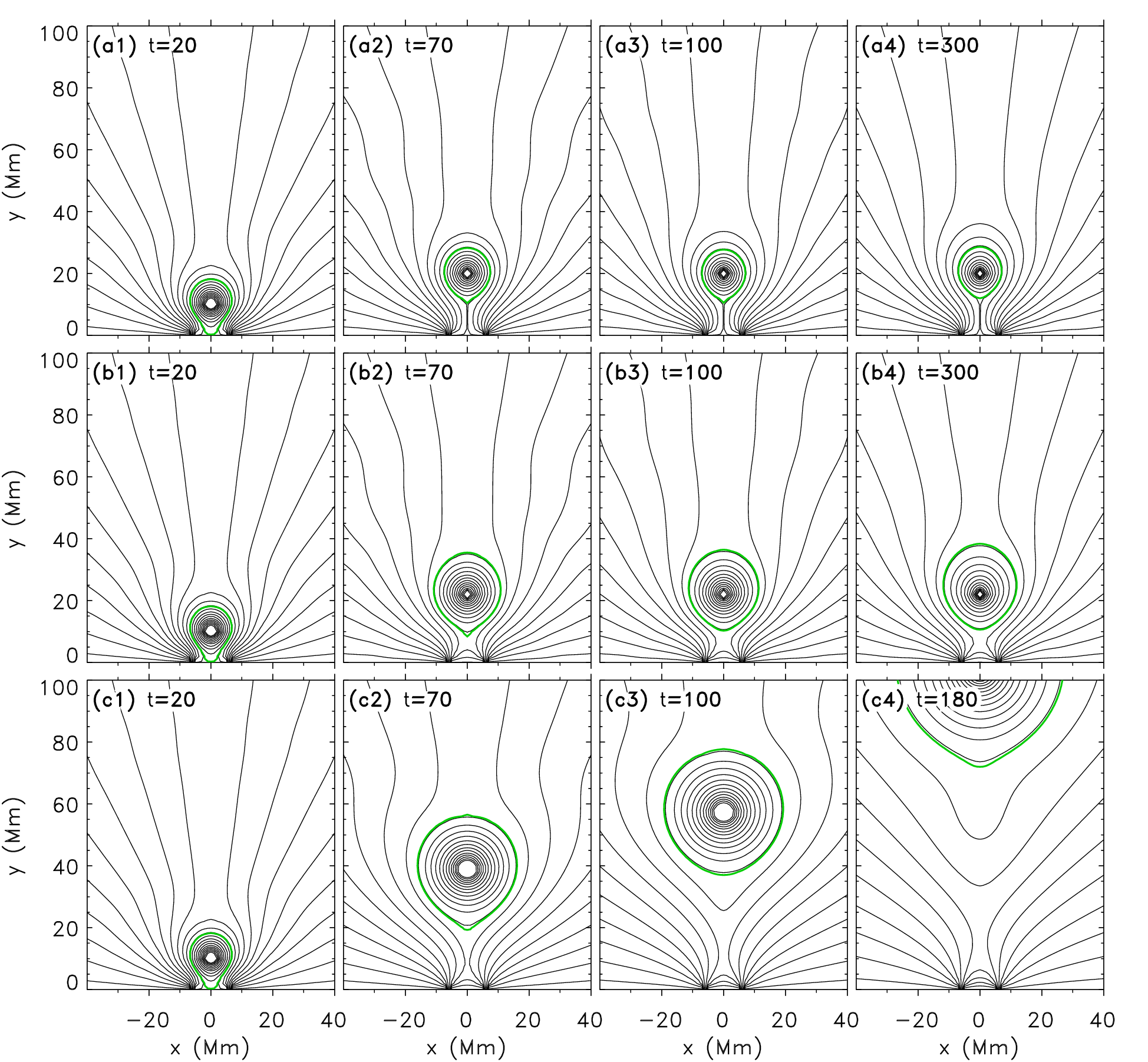}
\caption{The temporal evolutions of the flux rope for Case II. Panels (a1)-(a4) illustrate the evolution of the flux rope during the relaxation to simulate the equilibrium state in \fig{fig:weak}(g). Panels (b1)-(b4) illustrate the evolution of the flux rope during the relaxation to simulate the equilibrium state in \fig{fig:weak}(k), which is the equilibrium state right before $\Phi_z=\Phi_{zc}^w$ is reached under the resistive condition; Panels (b1)-(b4) illustrate the evolution right after $\Phi_z=\Phi_{zc}^w$ is reached under the resistive condition.}\label{fig:weak_evo}
\end{figure*}

The equilibrium manifold under the ideal MHD condition for Case II is plotted in \fig{fig:weak}(e), and all the equilibrium states have the same $\Phi_p=2.980\times10^{9}$ Mx cm$^{-1}$. Different from that in \fig{fig:weak}(a), the equilibrium manifold under the ideal MHD condition for Case II is continuous: the height of the rope axis, $H$, gradually rises as $\Phi_z$ increases, and the magnetic configurations of some equilibrium states are illustrated in \fig{fig:weak}(f)-\ref{fig:weak}(h). The evolution of the flux rope system during the relaxation to simulate the equilibrium state in \fig{fig:weak}(g) is shown in \fig{fig:weak_evo}(a1)-\ref{fig:weak_evo}(a4). After $\Phi_z$ is increased to $1.613\times10^{18}$ Mx at $t=20\tau_A$ (\fig{fig:weak_evo}(a1)), the flux rope gradually rises and then is suspended in the corona (\fig{fig:weak_evo}(a2)-\ref{fig:weak_evo}(a3)), until it reaches equilibrium (\fig{fig:weak_evo}(a4), the same as \fig{fig:weak}(g)). Although a current sheet is formed by the magnetic field of opposite directions below the rope, the magnetic reconnection is prohibited by the simulating procedures introduced in \sect{sec:method}. These simulation results demonstrate that the flux rope system in Case II is non-catastrophic so that no eruption could be initiated under the ideal MHD condition. As introduced in \sect{sec:initial}, the only difference in the characteristic parameters of the background field in Case I and Case II is $\psi_c$: the intensity of the background magnetic field originating from the photospheric source regions in Case II is much weaker than those in Case I, i.e. Case II is a ``weak'' case. This result is consistent with the conclusions in \cite{Zhang2017a,Zhang2017}. 
\par
To investigate the influence of magnetic reconnection on the catastrophic behaviors of flux rope systems with weak photospheric source regions, we also simulate the equilibrium manifold under the resistive condition for Case II, as plotted in \fig{fig:weak}(i). Obviously, if there is magnetic reconnection in the flux rope system, the equilibrium manifold is discontinuous: there is a critical axial magnetic flux $\Phi_{zc}^w\approx1.617\times10^{18}$ Mx, which is marked by the red vertical dotted line in \fig{fig:weak}(i). As shown in \fig{fig:weak}(i), $H$ gradually rises as $\Phi_z$ increases before reaching $\Phi_{zc}^w$. \fig{fig:weak}(k) illustrate the magnetic configuration of the equilibrium state with $\Phi_z=1.613\times10^{18}$ Mx, i.e., the equilibrium state right before $\Phi_{zc}^w$ is reached, and the evolution of the flux rope system to reach this equilibrium state is shown in \fig{fig:weak_evo}(b1)-\ref{fig:weak_evo}(b4). Comparing with the simulation results without reconnection in \fig{fig:weak_evo}(a1)-\ref{fig:weak_evo}(a4), it is demonstrated that magnetic reconnection occurs in the current sheet below the flux rope (\fig{fig:weak_evo}(b2)-\ref{fig:weak_evo}(b3)), resulting in closed arcades below the rope. The flux rope does not further rises but remains suspended, and eventually evolves to equilibrium (\fig{fig:weak_evo}(b4), the same as \fig{fig:weak}(g)). As introduced in \sect{sec:initial}, $B_z=0$ in the background field, so that the reconnection has no effect on the total axial magnetic flux of the rope. If $\Phi_z$ increases to reach $\Phi_{zc}^w$, the flux rope is initiated to erupt, as shown by the evolution of the flux rope system illustrated in \fig{fig:weak_evo}(c1)-\ref{fig:weak_evo}(c4): the flux rope does not remains suspended in the corona, but keeps rising after $\Phi_z$ is increased to $\Phi_{zc}^w$, resulting in the eruption of the flux rope. Obviously, the evolutions of the flux rope system before and after $\Phi_z$ reaches $\Phi_{zc}^w$ are quite different if magnetic reconnection is included, indicating that there is a catastrophe in the flux rope system, and $\Phi_{zc}^w$ is the catastrophic point. We note that there is no equilibrium state if the catastrophe occurs under the ressitive condition, so that we illustrate the state at $t=180\tau_A$ in \fig{fig:weak}(l) (the same as \fig{fig:weak_evo}(c4)) as the characteristic state for $\Phi_z=\Phi_{zc}^w$, and the corresponding red point in \fig{fig:weak}(i) is plotted at the top boundary. Our simulation results suggest that the magnetic flux rope system with weak photospheric source regions could also be catastrophic if magnetic reconnection is included in the system.
\par

\subsection{Compact cases}
\label{sec:compact}
\begin{figure*}
\includegraphics[width=\hsize]{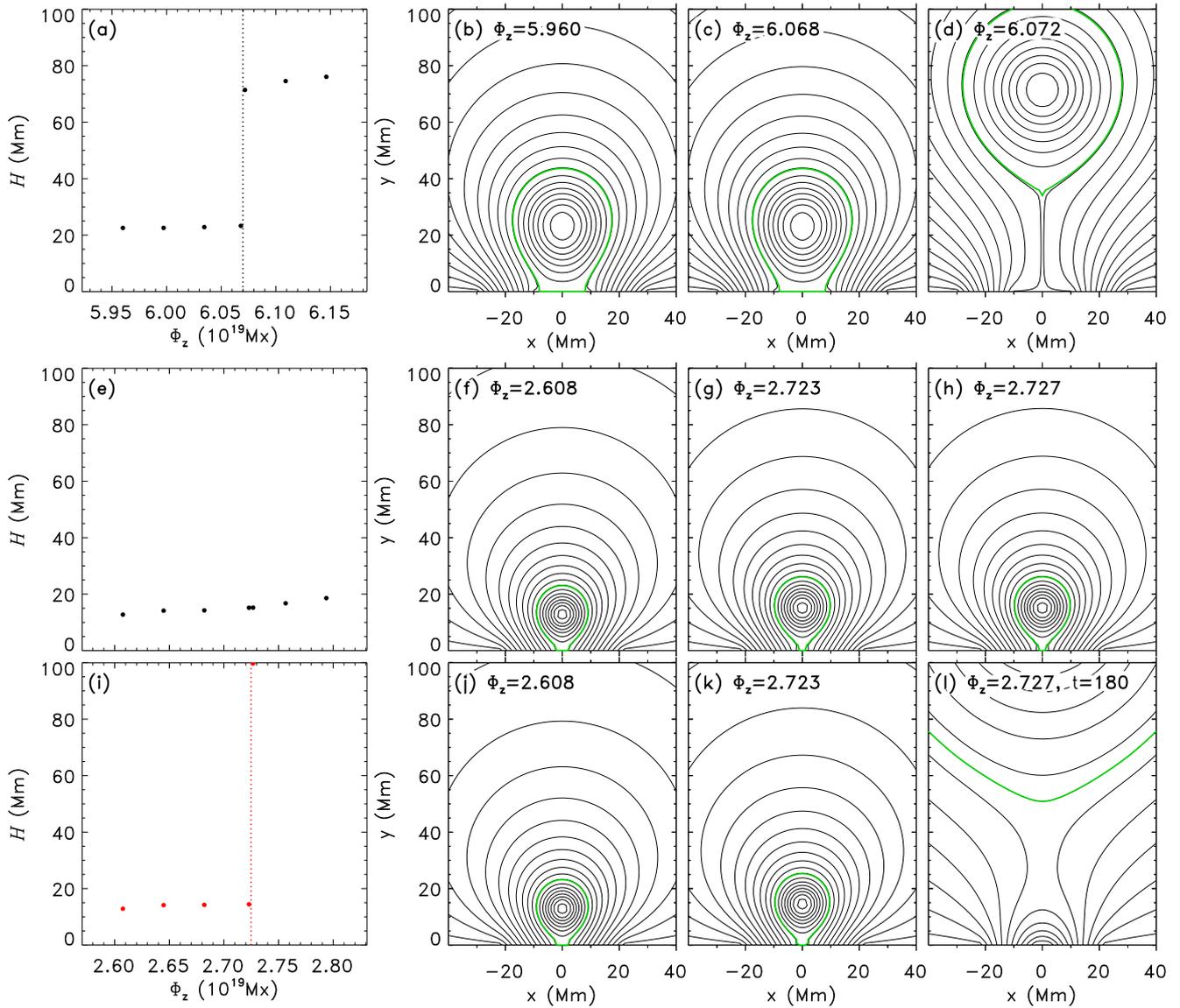}
\caption{The simulation results for Case III and Case IV. Panel (a)-(d) are the simulation results for Case III under the ideal MHD condition. Panels (e)-(h) and panels (i)-(l) are the simulation results for Case IV under the ideal MHD condition and the resistive condition, respectively; the red vertical dotted line in panel (i) denote the catastrophic point $\Phi_{zc}^c$. The green curves mark the boundary of the flux rope.}\label{fig:compact}
\end{figure*}

\begin{figure*}
\centering
\includegraphics[width=0.8\hsize]{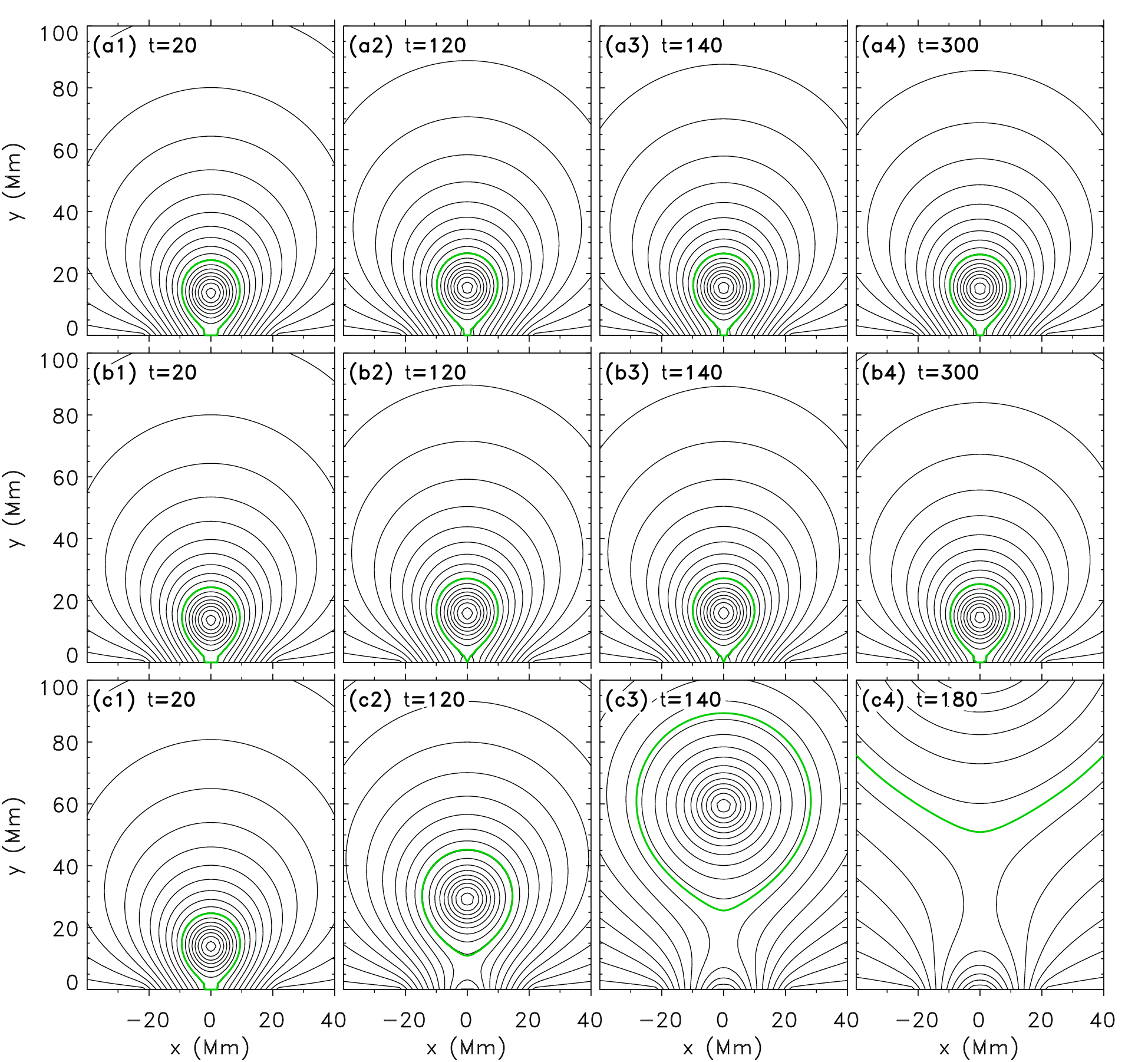}
\caption{The temporal evolution of the flux rope for Case IV. Panels (a1)-(a4) illustrate the evolution of the flux rope during the relaxation to simulate the equilibrium state in \fig{fig:compact}(g). Panels (b1)-(b4) illustrate the evolution of the flux rope during the relaxation to simulate the equilibrium state in \fig{fig:compact}(k), which is the equilibrium state right before $\Phi_z=\Phi_{zc}^c$ is reached under the resistive condition; Panels (b1)-(b4) illustrate the evolution right after $\Phi_z=\Phi_{zc}^c$ is reached under the resistive condition.}\label{fig:compact_evo}
\end{figure*}

The simulation results under the ideal MHD condition for Case III and Case IV are illustrated in \fig{fig:compact}(a)-\ref{fig:compact}(d) and \fig{fig:compact}(e)-\ref{fig:compact}(h), respectively, in which all the simulated equilibrium states have the same $\Phi_p=1.490\times10^{10}$ Mx cm$^{-1}$. As introduced in \sect{sec:initial} and \tbl{tbl:para}, the distance between the photospheric source regions is much smaller in Case IV than that in Case III. As shown in \fig{fig:compact}(a), the flux rope system in Case III is catastrophic: the catastrophe occurs if $\Phi_z$ reaches $6.072\times10^{19}$ Mx, as marked by the black vertical dotted line in \fig{fig:compact}(a). The flux rope system in Case IV, however, is non-catastrophic; its equilibrium manifold is continuous. This indicates Case IV is a ``compact'' case, which is also consistent with \cite{Zhang2017a,Zhang2017}. The evolution of the flux rope system to reach the equilibrium state in \fig{fig:compact}(g) is shown in \fig{fig:compact_evo}(a1)-\ref{fig:compact_evo}(a4).
\par
Similar as those in \sect{sec:weak}, we also simulate the equilibrium manifold under the ressitive condition for Case IV, so as to investigate the influence of magnetic reconnection on the catastrophic behaviors of flux rope systems with compact photospheric source regions. As plotted in \fig{fig:compact}(i), there is a critical axial magnetic flux $\Phi_{zc}^c\approx2.727\times10^{19}$ Mx if magnetic reconnection is included in the simulation, as marked by the red vertical dotted lines in \fig{fig:compact}(i). The evolutions of the flux rope system before and after $\Phi_z$ reaches $\Phi_{zc}^c$ are quite different: if $\Phi_z$ is smaller than $\Phi_{zc}^c$, the flux rope does not erupt but evolves to equilibrium (\fig{fig:compact_evo}(b1)-\ref{fig:compact_evo}(b4)); if $\Phi_{zc}^c$ is reached, the flux rope erupts(\fig{fig:compact_evo}(c1)-\ref{fig:compact_evo}(c4)). This indicates that a catastrophe associated with the increasing axial magnetic flux of the rope could occur under the ressitive condition. For comparison, the topologies of the equilibrium states before and after $\Phi_z$ reaches $\Phi_{zc}^c$ are quite similar if there is no reconnection (\fig{fig:compact}(g) and \fig{fig:compact}(h)). This further confirms that it is the magnetic reconnection that changes the catastrophic behaviors of the flux rope system versus varying axial flux. Therefore, the flux rope system with compact photospheric source regions could also be catastrophic if there is magnetic reconnection in the rope system.

\section{Discussion and Conclusion}
\label{sec:dc}
In this paper, we investigate the influence of magnetic reconnection on the catastrophic behaviors of coronal flux rope systems associated with increasing axial magnetic flux of the rope. Under the ideal MHD condition, our simulation results demonstrate that coronal flux rope systems with either too compact or too weak photospheric source regions are non-catastrophic versus varying axial magnetic flux. Under the resistive condition, however, both the flux rope system with too compact and too weak photospheric source regions could change to be catastrophic: the catastrophe occurs when the axial flux of the rope increases to reach the critical value, so that the eruption of the flux rope is initiated.
\par
Our simulation results demonstrate that the equilibrium manifolds of a coronal flux rope system could be quite different under the ideal MHD and the resistive condition (e.g., \fig{fig:weak}(e) and \fig{fig:weak}(i)). This indicates that magnetic reconnection should play a considerable role in determining the force balance and the corresponding the loss of equilibrium associated with increasing axial magnetic flux in coronal flux rope systems. We note that the continuous equilibrium manifold (\fig{fig:weak}(e)) is different from the discontinuous equilibrium manifold (\fig{fig:weak}(i)) in essence. It is widely accepted that solar eruptions should be caused by photospheric and coronal disturbances \citep{Lin2003a}, the spatial range of which, however, is much smaller than that of solar eruptions \citep{Priest2014a}. If the equilibrium manifold of a flux rope system is continuous, the variations of $\Phi_z$ caused by photospheric or coronal disturbances could only cause flux rope movements on a spatial scale comparable to the disturbance itself. In contrast, if the equilibrium manifold is discontinuous, there is a critical axial flux of $\Phi_z$, so that even an infinitesimal increment of $\Phi_z$ to reach this critical value could cause the catastrophe of the flux rope, resulting in much larger scale of movement of the flux rope than that of the disturbances. Therefore, solar eruptions could only be initiated in the coronal flux rope system whose equilibrium manifold is discontinuous.
\par
In previous studies, the forecasting of solar eruptions is generally based on the photospheric magnetic flux distributions within solar active regions \cite[e.g.,][]{Leka2007a,Bobra2015,Toriumi2019,Korsos2020}. However, as revealed by our simulation results, not only do different photospheric flux distributions result in different evolutions of the flux rope system, but the catastrophic behaviors of a particular flux rope system with given photospheric flux distribution could also differ under the ideal MHD condition and the resistive condition. This implies that two active regions with similar photospheric flux distribution might be quite different in their ability to initiate solar eruptions, provided that the local resistivity and characteristic spatial scale differs in the two active regions. Therefore, it should not be sufficient to predict whether a solar active region will be a potential source region for solar eruptions based solely on its photospheric magnetic flux distributions. Our simulation results suggest that both the magnetic topology and the local physical parameters related to magnetic reconnection determine whether increasing axial flux is able to cause flux rope eruptions. 
\par
There are also catastrophes associated with other physical parameters of flux rope systems, such as increasing shear of the background field \citep{Chen2006a} and decreasing mass of the rope \citep{Jenkins2019,Zhang2021}. It could be inferred that the catastrophic behaviors associated with those parameters might probably also be determined by both the magnetic topology and whether the magnetic system is resistive. In our future work, we will carry out more simulations to further investigate the influence of magnetic reconnection on different kinds catastrophes.

\section*{Conflict of Interest Statement}
The authors declare that the research was conducted in the absence of any commercial or financial relationships that could be construed as a potential conflict of interest.

\section*{Author Contributions}
Quanhao Zhang carried out the numerical simulation, analyzed the simulation results, and prepared the manuscript. Xin Cheng contributed to the initial inspiration of this study, and the improvement of the numerical model. Rui Liu contributed to the analysis of the simulation results. Anchuan Song, Xiaolei Li, and Yuming Wang joined in the discussions about the simulation results.

\section*{Funding}
This research is supported by the National Natural Science Foundation of China (NSFC 42174213, 42188101, 42130204, 41804161, 41774178, 41774150), the Strategic Priority Program of CAS (XDB41000000 and XDA15017300), the Informatization Plan of Chinese Academy of Sciences, Grant No.CAS-WX2021PY-0101, and the Key Research Program of the Chinese Academy of Sciences, Grant NO. ZDBS-SSW-TLC00103. This research is also supported by USTC Research Funds of the Double First-Class Initiative.

\section*{Acknowledgments}
The authors acknowledge for the support from National Space Science Data Center, National Science \verb"&" Technology Infrastructure of China (www.nssdc.ac.cn). Quanhao Zhang acknowledge for the support from Young Elite Scientist Sponsorship Program by the China Association for Science and Technology (CAST).

\section*{Data Availability Statement}
The simulation results in this study are included in the article; if there is any further inquiry, please contact the corresponding author.


\begin{thebibliography}{70}
\providecommand{\natexlab}[1]{#1}
\expandafter\ifx\csname urlstyle\endcsname\relax
  \providecommand{\doi}[1]{doi:\discretionary{}{}{}#1}\else
  \providecommand{\doi}{doi:\discretionary{}{}{}\begingroup
  \urlstyle{rm}\Url}\fi
\providecommand{\selectlanguage}[1]{\relax}
\providecommand{\bibAnnoteFile}[1]{%
  \IfFileExists{#1}{\begin{quotation}\noindent\textsc{Key:} #1\\
  \textsc{Annotation:}\ \input{#1}\end{quotation}}{}}
\providecommand{\bibAnnote}[2]{%
  \begin{quotation}\noindent\textsc{Key:} #1\\
  \textsc{Annotation:}\ #2\end{quotation}}

\bibitem[{{Antiochos} et~al.(1999){Antiochos}, {DeVore}, and
  {Klimchuk}}]{Antiochos1999a}
{Antiochos}, S.~K., {DeVore}, C.~R., and {Klimchuk}, J.~A. (1999).
\newblock A model for solar coronal mass ejections.
\newblock \emph{\apj} 510, 485--493.
\newblock \doi{10.1086/306563}
\input{Antiochos1999a}

\bibitem[{Anzer and Heinzel(2007)}]{Anzer2007}
Anzer, U. and Heinzel, P. (2007).
\newblock Is the magnetic field in quiescent prominences force-free?
\newblock \emph{\aap} 467, 1285--1288.
\newblock \doi{10.1051/0004-6361:20066817}
\input{Anzer2007}

\bibitem[{{Archontis} and {Hood}(2008)}]{Archontis2008b}
{Archontis}, V. and {Hood}, A.~W. (2008).
\newblock {A Flux Emergence Model for Solar Eruptions}.
\newblock \emph{\apjl} 674, L113.
\newblock \doi{10.1086/529377}
\input{Archontis2008b}

\bibitem[{{Aulanier} et~al.(2010){Aulanier}, {T{\"o}r{\"o}k}, {D{\'e}moulin},
  and {DeLuca}}]{Aulanier2010a}
{Aulanier}, G., {T{\"o}r{\"o}k}, T., {D{\'e}moulin}, P., and {DeLuca}, E.~E.
  (2010).
\newblock Formation of torus-unstable flux ropes and electric currents in
  erupting sigmoids.
\newblock \emph{\apj} 708, 314--333.
\newblock \doi{10.1088/0004-637X/708/1/314}
\input{Aulanier2010a}

\bibitem[{Bemporad et~al.(2022)Bemporad, Andretta, Susino, Mancuso, Spadaro,
  Mierla et~al.}]{Bemporad2022}
Bemporad, A., Andretta, V., Susino, R., Mancuso, S., Spadaro, D., Mierla, M.,
  et~al. (2022).
\newblock Coronal mass ejection followed by a prominence eruption and a plasma
  blob as observed by solar orbiter.
\newblock \emph{\aap} 665, A7.
\newblock \doi{10.1051/0004-6361/202243162}
\input{Bemporad2022}

\bibitem[{Bian et~al.(2022)Bian, Jiang, and Feng}]{Bian2022}
Bian, X., Jiang, C., and Feng, X. (2022).
\newblock The role of photospheric converging motion in initiation of solar
  eruptions.
\newblock \emph{Frontiers in Astronomy and Space Sciences} 9, 982108.
\newblock \doi{10.3389/fspas.2022.982108}
\input{Bian2022}

\bibitem[{Bobra and Couvidat(2015)}]{Bobra2015}
Bobra, M.~G. and Couvidat, S. (2015).
\newblock Solar flare prediction using sdo/hmi vector magnetic field data with
  a machine-learning algorithm.
\newblock \emph{\apj} 798, 135.
\newblock \doi{10.1088/0004-637X/798/2/135}
\input{Bobra2015}

\bibitem[{{Bobra} et~al.(2008){Bobra}, {van Ballegooijen}, and
  {DeLuca}}]{Bobra2008}
{Bobra}, M.~G., {van Ballegooijen}, A.~A., and {DeLuca}, E.~E. (2008).
\newblock {Modeling Nonpotential Magnetic Fields in Solar Active Regions}.
\newblock \emph{\apj} 672, 1209--1220.
\newblock \doi{10.1086/523927}
\input{Bobra2008}

\bibitem[{Chen et~al.(2020)Chen, Yu, Reeves, and Gary}]{Chen2020a}
Chen, B., Yu, S., Reeves, K.~K., and Gary, D.~E. (2020).
\newblock Microwave spectral imaging of an erupting magnetic flux rope:
  Implications for the standard solar flare model in three dimensions.
\newblock \emph{\apjl} 895, L50.
\newblock \doi{10.3847/2041-8213/ab901a}
\input{Chen2020a}

\bibitem[{Chen et~al.(2015)Chen, Zhang, Ma, Yang, Li, Huang et~al.}]{Chen2015}
Chen, H., Zhang, J., Ma, S., Yang, S., Li, L., Huang, X., et~al. (2015).
\newblock Confined flares in solar active region 12192 from 2014 october 18 to
  29.
\newblock \emph{\apjl} 808, L24.
\newblock \doi{10.1088/2041-8205/808/1/L24}
\input{Chen2015}

\bibitem[{{Chen} and {Shibata}(2000)}]{Chen2000a}
{Chen}, P.~F. and {Shibata}, K. (2000).
\newblock An emerging flux trigger mechanism for coronal mass ejections.
\newblock \emph{\apj} 545, 524--531.
\newblock \doi{10.1086/317803}
\input{Chen2000a}

\bibitem[{Chen et~al.(2004)Chen, Shibata, Brooks, and Isobe}]{Chen2004}
Chen, P.~F., Shibata, K., Brooks, D.~H., and Isobe, H. (2004).
\newblock A reexamination of the evidence for reconnection inflow.
\newblock \emph{\apjl} 602, L61--L64.
\newblock \doi{10.1086/382479}
\input{Chen2004}

\bibitem[{{Chen} et~al.(2006){Chen}, {Chen}, and {Hu}}]{Chen2006a}
{Chen}, Y., {Chen}, X.~H., and {Hu}, Y.~Q. (2006).
\newblock Catastrophe of coronal flux rope in unsheared and sheared bipolar
  magnetic fields.
\newblock \emph{\apj} 644, 587--591.
\newblock \doi{10.1086/503540}
\input{Chen2006a}

\bibitem[{{Chen} et~al.(2007){Chen}, {Hu}, and {Sun}}]{Chen2007a}
{Chen}, Y., {Hu}, Y.~Q., and {Sun}, S.~J. (2007).
\newblock Catastrophic eruption of magnetic flux rope in the corona and solar
  wind with and without magnetic reconnection.
\newblock \emph{\apj} 665, 1421--1427.
\newblock \doi{10.1086/519551}
\input{Chen2007a}

\bibitem[{{Cheng} et~al.(2014){Cheng}, {Ding}, {Zhang}, {Sun}, {Guo}, {Wang}
  et~al.}]{Cheng2014}
{Cheng}, X., {Ding}, M.~D., {Zhang}, J., {Sun}, X.~D., {Guo}, Y., {Wang},
  Y.~M., et~al. (2014).
\newblock {Formation of a Double-decker Magnetic Flux Rope in the Sigmoidal
  Solar Active Region 11520}.
\newblock \emph{\apj} 789, 93.
\newblock \doi{10.1088/0004-637X/789/2/93}
\input{Cheng2014}

\bibitem[{Cheng et~al.(2020)Cheng, Zhang, Kliem, T{\"o}r{\"o}k, Xing, Zhou
  et~al.}]{Cheng2020a}
Cheng, X., Zhang, J., Kliem, B., T{\"o}r{\"o}k, T., Xing, C., Zhou, Z.~J.,
  et~al. (2020).
\newblock Initiation and early kinematic evolution of solar eruptions.
\newblock \emph{\apj} 894, 85.
\newblock \doi{10.3847/1538-4357/ab886a}
\input{Cheng2020a}

\bibitem[{Cheung et~al.(2019)Cheung, Rempel, Chintzoglou, Chen, Testa,
  Mart{\'\i}nez-Sykora et~al.}]{Cheung2019}
Cheung, M. C.~M., Rempel, M., Chintzoglou, G., Chen, F., Testa, P.,
  Mart{\'\i}nez-Sykora, J., et~al. (2019).
\newblock A comprehensive three-dimensional radiative magnetohydrodynamic
  simulation of a solar flare.
\newblock \emph{Nature Astronomy} 3, 160--166.
\newblock \doi{10.1038/s41550-018-0629-3}
\input{Cheung2019}

\bibitem[{{D{\'e}moulin} and {Aulanier}(2010)}]{Demoulin2010a}
{D{\'e}moulin}, P. and {Aulanier}, G. (2010).
\newblock Criteria for flux rope eruption: Non-equilibrium versus torus
  instability.
\newblock \emph{\apj} 718, 1388--1399.
\newblock \doi{10.1088/0004-637X/718/2/1388}
\input{Demoulin2010a}

\bibitem[{Devi et~al.(2021)Devi, D{\'e}moulin, Chandra, Joshi, Schmieder, and
  Joshi}]{Devi2021}
Devi, P., D{\'e}moulin, P., Chandra, R., Joshi, R., Schmieder, B., and Joshi,
  B. (2021).
\newblock Observations of a prominence eruption and loop contraction.
\newblock \emph{\aap} 647, A85.
\newblock \doi{10.1051/0004-6361/202040042}
\input{Devi2021}

\bibitem[{{Fan}(2020)}]{Fan2020}
{Fan}, Y. (2020).
\newblock {Simulations of Prominence Eruption Preceded by Large-amplitude
  Longitudinal Oscillations and Draining}.
\newblock \emph{\apj} 898, 34.
\newblock \doi{10.3847/1538-4357/ab9d7f}
\input{Fan2020}

\bibitem[{{Forbes}(1990)}]{Forbes1990a}
{Forbes}, T.~G. (1990).
\newblock Numerical simulation of a catastrophe model for coronal mass
  ejections.
\newblock \emph{\jgr} 95, 11919--11931.
\newblock \doi{10.1029/JA095iA08p11919}
\input{Forbes1990a}

\bibitem[{{Forbes} and {Priest}(1995)}]{Forbes1995a}
{Forbes}, T.~G. and {Priest}, E.~R. (1995).
\newblock Photospheric magnetic field evolution and eruptive flares.
\newblock \emph{\apj} 446, 377.
\newblock \doi{10.1086/175797}
\input{Forbes1995a}

\bibitem[{{Gopalswamy} et~al.(2018){Gopalswamy}, {Akiyama}, {Yashiro}, and
  {Xie}}]{Gopalswamy2018a}
{Gopalswamy}, N., {Akiyama}, S., {Yashiro}, S., and {Xie}, H. (2018).
\newblock {Coronal flux ropes and their interplanetary counterparts}.
\newblock \emph{Journal of Atmospheric and Solar-Terrestrial Physics} 180,
  35--45.
\newblock \doi{10.1016/j.jastp.2017.06.004}
\input{Gopalswamy2018a}

\bibitem[{{Guo} et~al.(2010){Guo}, {Ding}, {Schmieder}, {Li}, {T{\"o}r{\"o}k},
  and {Wiegelmann}}]{Guo2010}
{Guo}, Y., {Ding}, M.~D., {Schmieder}, B., {Li}, H., {T{\"o}r{\"o}k}, T., and
  {Wiegelmann}, T. (2010).
\newblock {Driving Mechanism and Onset Condition of a Confined Eruption}.
\newblock \emph{\apjl} 725, L38--L42.
\newblock \doi{10.1088/2041-8205/725/1/L38}
\input{Guo2010}

\bibitem[{Hillier et~al.(2012)Hillier, Hillier, and Tripathi}]{Hillier2012}
Hillier, A., Hillier, R., and Tripathi, D. (2012).
\newblock Determination of prominence plasma {\ensuremath{\beta}} from the
  dynamics of rising plumes.
\newblock \emph{\apj} 761, 106.
\newblock \doi{10.1088/0004-637X/761/2/106}
\input{Hillier2012}

\bibitem[{Hu and Liu(2000)}]{Hu2000a}
Hu, Y.~Q. and Liu, W. (2000).
\newblock A 2.5-dimensional ideal magnetohydrodynamic model for coronal
  magnetic flux ropes.
\newblock \emph{\apj} 540, 1119--1125.
\newblock \doi{10.1086/309381}
\input{Hu2000a}

\bibitem[{{Inoue} et~al.(2015){Inoue}, {Hayashi}, {Magara}, {Choe}, and
  {Park}}]{Inoue2015}
{Inoue}, S., {Hayashi}, K., {Magara}, T., {Choe}, G.~S., and {Park}, Y.~D.
  (2015).
\newblock {Magnetohydrodynamic Simulation of the X2.2 Solar Flare on 2011
  February 15. II. Dynamics Connecting the Solar Flare and the Coronal Mass
  Ejection}.
\newblock \emph{\apj} 803, 73.
\newblock \doi{10.1088/0004-637X/803/2/73}
\input{Inoue2015}

\bibitem[{{Isenberg} and {Forbes}(2007)}]{Isenberg2007a}
{Isenberg}, P.~A. and {Forbes}, T.~G. (2007).
\newblock A three-dimensional line-tied magnetic field model for solar
  eruptions.
\newblock \emph{\apj} 670, 1453--1466.
\newblock \doi{10.1086/522025}
\input{Isenberg2007a}

\bibitem[{{Isenberg} et~al.(1993){Isenberg}, {Forbes}, and
  {Demoulin}}]{Isenberg1993a}
{Isenberg}, P.~A., {Forbes}, T.~G., and {Demoulin}, P. (1993).
\newblock Catastrophic evolution of a force-free flux rope: A model for
  eruptive flares.
\newblock \emph{\apj} 417, 368.
\newblock \doi{10.1086/173319}
\input{Isenberg1993a}

\bibitem[{Jenkins et~al.(2019)Jenkins, Hopwood, D{\'e}moulin, Valori, Aulanier,
  Long et~al.}]{Jenkins2019}
Jenkins, J.~M., Hopwood, M., D{\'e}moulin, P., Valori, G., Aulanier, G., Long,
  D.~M., et~al. (2019).
\newblock Modeling the effect of mass-draining on prominence eruptions.
\newblock \emph{\apj} 873, 49.
\newblock \doi{10.3847/1538-4357/ab037a}
\input{Jenkins2019}

\bibitem[{Jenkins et~al.(2018)Jenkins, Long, van Driel-Gesztelyi, and
  Carlyle}]{Jenkins2018}
Jenkins, J.~M., Long, D.~M., van Driel-Gesztelyi, L., and Carlyle, J. (2018).
\newblock Understanding the role of mass-unloading in a filament eruption.
\newblock \emph{\solphys} 293, 7.
\newblock \doi{10.1007/s11207-017-1224-y}
\input{Jenkins2018}

\bibitem[{Jiang et~al.(2021)Jiang, Feng, Liu, Yan, Hu, Moore
  et~al.}]{Jiang_2021}
Jiang, C., Feng, X., Liu, R., Yan, X., Hu, Q., Moore, R.~L., et~al. (2021).
\newblock A fundamental mechanism of solar eruption initiation.
\newblock \emph{Nature Astronomy} 5, 1126--1138.
\newblock \doi{10.1038/s41550-021-01414-z}
\input{Jiang_2021}

\bibitem[{Jiang et~al.(2016)Jiang, Wu, Yurchyshyn, Wang, Feng, and
  Hu}]{Jiang2016a}
Jiang, C., Wu, S.~T., Yurchyshyn, V., Wang, H., Feng, X., and Hu, Q. (2016).
\newblock How did a major confined flare occur in super solar active region
  12192?
\newblock \emph{\apj} 828, 62.
\newblock \doi{10.3847/0004-637X/828/1/62}
\input{Jiang2016a}

\bibitem[{Keppens et~al.(2019)Keppens, Guo, Makwana, Mei, Ripperda, Xia
  et~al.}]{Keppens2019}
Keppens, R., Guo, Y., Makwana, K., Mei, Z., Ripperda, B., Xia, C., et~al.
  (2019).
\newblock Ideal mhd instabilities for coronal mass ejections: interacting
  current channels and particle acceleration.
\newblock \emph{Reviews of Modern Plasma Physics} 3, 14.
\newblock \doi{10.1007/s41614-019-0035-z}
\input{Keppens2019}

\bibitem[{{Kliem} et~al.(2014){Kliem}, {Lin}, {Forbes}, {Priest}, and
  {T{\"o}r{\"o}k}}]{Kliem2014}
{Kliem}, B., {Lin}, J., {Forbes}, T.~G., {Priest}, E.~R., and {T{\"o}r{\"o}k},
  T. (2014).
\newblock Catastrophe versus instability for the eruption of a toroidal solar
  magnetic flux rope.
\newblock \emph{\apj} 789, 46.
\newblock \doi{10.1088/0004-637X/789/1/46}
\input{Kliem2014}

\bibitem[{Kors{\'o}s et~al.(2020)Kors{\'o}s, Georgoulis, Gyenge, Bisoi, Yu,
  Poedts et~al.}]{Korsos2020}
Kors{\'o}s, M.~B., Georgoulis, M.~K., Gyenge, N., Bisoi, S.~K., Yu, S., Poedts,
  S., et~al. (2020).
\newblock Solar flare prediction using magnetic field diagnostics above the
  photosphere.
\newblock \emph{\apj} 896, 119.
\newblock \doi{10.3847/1538-4357/ab8fa2}
\input{Korsos2020}

\bibitem[{Lamy et~al.(2019)Lamy, Floyd, Boclet, Wojak, Gilardy, and
  Barlyaeva}]{Lamy2019}
Lamy, P.~L., Floyd, O., Boclet, B., Wojak, J., Gilardy, H., and Barlyaeva, T.
  (2019).
\newblock Coronal mass ejections over solar cycles 23 and 24.
\newblock \emph{\ssr} 215, 39.
\newblock \doi{10.1007/s11214-019-0605-y}
\input{Lamy2019}

\bibitem[{{Leka} and {Barnes}(2007)}]{Leka2007a}
{Leka}, K.~D. and {Barnes}, G. (2007).
\newblock Photospheric magnetic field properties of flaring versus flare-quiet
  active regions. iv. a statistically significant sample.
\newblock \emph{\apj} 656, 1173--1186.
\newblock \doi{10.1086/510282}
\input{Leka2007a}

\bibitem[{{Li} et~al.(2016{\natexlab{a}}){Li}, {Liu}, {Elmhamdi}, and
  {Kordi}}]{Li2016}
{Li}, H., {Liu}, Y., {Elmhamdi}, A., and {Kordi}, A.-S. (2016{\natexlab{a}}).
\newblock {Relationship between Distribution of Magnetic Decay Index and
  Filament Eruptions}.
\newblock \emph{\apj} 830, 132.
\newblock \doi{10.3847/0004-637X/830/2/132}
\input{Li2016}

\bibitem[{{Li} et~al.(2016{\natexlab{b}}){Li}, {Yang}, {Hou}, and
  {Zhang}}]{Li2016a}
{Li}, T., {Yang}, K., {Hou}, Y., and {Zhang}, J. (2016{\natexlab{b}}).
\newblock {Slipping Magnetic Reconnection of Flux-rope Structures as a
  Precursor to an Eruptive X-class Solar Flare}.
\newblock \emph{\apj} 830, 152.
\newblock \doi{10.3847/0004-637X/830/2/152}
\input{Li2016a}

\bibitem[{{Lin} and {Forbes}(2000)}]{Lin2000a}
{Lin}, J. and {Forbes}, T.~G. (2000).
\newblock Effects of reconnection on the coronal mass ejection process.
\newblock \emph{\jgr} 105, 2375--2392.
\newblock \doi{10.1029/1999JA900477}
\input{Lin2000a}

\bibitem[{{Lin} et~al.(2003){Lin}, {Soon}, and {Baliunas}}]{Lin2003a}
{Lin}, J., {Soon}, W., and {Baliunas}, S.~L. (2003).
\newblock Theories of solar eruptions: a review.
\newblock \emph{\nar} 47, 53--84.
\newblock \doi{10.1016/S1387-6473(02)00271-3}
\input{Lin2003a}

\bibitem[{{Lin} and {van Ballegooijen}(2002)}]{Lin2002b}
{Lin}, J. and {van Ballegooijen}, A.~A. (2002).
\newblock Catastrophic and noncatastrophic mechanisms for coronal mass
  ejections.
\newblock \emph{\apj} 576, 485--492.
\newblock \doi{10.1086/341737}
\input{Lin2002b}

\bibitem[{Liu(2020)}]{Liu2020a}
Liu, R. (2020).
\newblock Magnetic flux ropes in the solar corona: structure and evolution
  toward eruption.
\newblock \emph{Research in Astronomy and Astrophysics} 20, 165.
\newblock \doi{10.1088/1674-4527/20/10/165}
\input{Liu2020a}

\bibitem[{{Longcope} and {Forbes}(2014)}]{Longcope2014a}
{Longcope}, D.~W. and {Forbes}, T.~G. (2014).
\newblock Breakout and tether-cutting eruption models are both catastrophic
  (sometimes).
\newblock \emph{\solphys} 289, 2091--2122.
\newblock \doi{10.1007/s11207-013-0464-8}
\input{Longcope2014a}

\bibitem[{Lugaz et~al.(2017)Lugaz, Farrugia, Winslow, Small, Manion, and
  Savani}]{Lugaz2017}
Lugaz, N., Farrugia, C.~J., Winslow, R.~M., Small, C.~R., Manion, T., and
  Savani, N.~P. (2017).
\newblock Importance of cme radial expansion on the ability of slow cmes to
  drive shocks.
\newblock \emph{\apj} 848, 75.
\newblock \doi{10.3847/1538-4357/aa8ef9}
\input{Lugaz2017}

\bibitem[{{Moore} et~al.(2001){Moore}, {Sterling}, {Hudson}, and
  {Lemen}}]{Moore2001a}
{Moore}, R.~L., {Sterling}, A.~C., {Hudson}, H.~S., and {Lemen}, J.~R. (2001).
\newblock Onset of the magnetic explosion in solar flares and coronal mass
  ejections.
\newblock \emph{\apj} 552, 833--848.
\newblock \doi{10.1086/320559}
\input{Moore2001a}

\bibitem[{{Priest}(2014)}]{Priest2014a}
{Priest}, E. (2014).
\newblock \emph{Magnetohydrodynamics of the Sun} (Cambridge University Press)
\input{Priest2014a}

\bibitem[{{Priest} and {Forbes}(1990)}]{Priest1990a}
{Priest}, E.~R. and {Forbes}, T.~G. (1990).
\newblock Magnetic field evolution during prominence eruptions and two-ribbon
  flares.
\newblock \emph{\solphys} 126, 319--350.
\newblock \doi{10.1007/BF00153054}
\input{Priest1990a}

\bibitem[{{Savcheva} et~al.(2012){Savcheva}, {van Ballegooijen}, and
  {DeLuca}}]{Savcheva2012b}
{Savcheva}, A.~S., {van Ballegooijen}, A.~A., and {DeLuca}, E.~E. (2012).
\newblock Field topology analysis of a long-lasting coronal sigmoid.
\newblock \emph{\apj} 744, 78.
\newblock \doi{10.1088/0004-637X/744/1/78}
\input{Savcheva2012b}

\bibitem[{{Shen} et~al.(2014){Shen}, {Shen}, {Zhang}, {Hess}, {Wang}, {Feng}
  et~al.}]{Shen2014}
{Shen}, F., {Shen}, C., {Zhang}, J., {Hess}, P., {Wang}, Y., {Feng}, X., et~al.
  (2014).
\newblock {Evolution of the 12 July 2012 CME from the Sun to the Earth:
  Data-constrained three-dimensional MHD simulations}.
\newblock \emph{Journal of Geophysical Research (Space Physics)} 119,
  7128--7141.
\newblock \doi{10.1002/2014JA020365}
\input{Shen2014}

\bibitem[{{Sterling} and {Moore}(2004)}]{Sterling2004}
{Sterling}, A.~C. and {Moore}, R.~L. (2004).
\newblock {Evidence for Gradual External Reconnection before Explosive Eruption
  of a Solar Filament}.
\newblock \emph{\apj} 602, 1024--1036.
\newblock \doi{10.1086/379763}
\input{Sterling2004}

\bibitem[{{Su} et~al.(2011){Su}, {Surges}, {van Ballegooijen}, {DeLuca}, and
  {Golub}}]{Su2011a}
{Su}, Y., {Surges}, V., {van Ballegooijen}, A., {DeLuca}, E., and {Golub}, L.
  (2011).
\newblock Observations and magnetic field modeling of the flare/coronal mass
  ejection event on 2010 april 8.
\newblock \emph{\apj} 734, 53.
\newblock \doi{10.1088/0004-637X/734/1/53}
\input{Su2011a}

\bibitem[{{Su} et~al.(2009){Su}, {van Ballegooijen}, {Lites}, {Deluca},
  {Golub}, {Grigis} et~al.}]{Su2009}
{Su}, Y., {van Ballegooijen}, A., {Lites}, B.~W., {Deluca}, E.~E., {Golub}, L.,
  {Grigis}, P.~C., et~al. (2009).
\newblock {Observations and Nonlinear Force-Free Field Modeling of Active
  Region 10953}.
\newblock \emph{\apj} 691, 105--114.
\newblock \doi{10.1088/0004-637X/691/1/105}
\input{Su2009}

\bibitem[{{Sun} et~al.(2007){Sun}, {Hu}, and {Chen}}]{Sun2007a}
{Sun}, S.~J., {Hu}, Y.~Q., and {Chen}, Y. (2007).
\newblock Influence of photospheric magnetic flux distribution on coronal flux
  rope catastrophe.
\newblock \emph{\apjl} 654, L167--L170.
\newblock \doi{10.1086/511304}
\input{Sun2007a}

\bibitem[{Toriumi and Wang(2019)}]{Toriumi2019}
Toriumi, S. and Wang, H. (2019).
\newblock Flare-productive active regions.
\newblock \emph{Living Reviews in Solar Physics} 16, 3.
\newblock \doi{10.1007/s41116-019-0019-7}
\input{Toriumi2019}

\bibitem[{{T{\"o}r{\"o}k} and {Kliem}(2003)}]{Torok2003a}
{T{\"o}r{\"o}k}, T. and {Kliem}, B. (2003).
\newblock The evolution of twisting coronal magnetic flux tubes.
\newblock \emph{\aap} 406, 1043--1059.
\newblock \doi{10.1051/0004-6361:20030692}
\input{Torok2003a}

\bibitem[{T{\"o}r{\"o}k et~al.(2013)T{\"o}r{\"o}k, Temmer, Valori, Veronig, van
  Driel-Gesztelyi, and Vr{\v{s}}nak}]{Toeroek2013}
T{\"o}r{\"o}k, T., Temmer, M., Valori, G., Veronig, A.~M., van Driel-Gesztelyi,
  L., and Vr{\v{s}}nak, B. (2013).
\newblock Initiation of coronal mass ejections by sunspot rotation.
\newblock \emph{\solphys} 286, 453--477.
\newblock \doi{10.1007/s11207-013-0269-9}
\input{Toeroek2013}

\bibitem[{{{\v S}vestka}(2001)}]{svestka2001a}
{{\v S}vestka}, Z. (2001).
\newblock Varieties of coronal mass ejections and their relation to flares.
\newblock \emph{\ssr} 95, 135--146.
\newblock \doi{10.1023/A:1005225208925}
\input{svestka2001a}

\bibitem[{{Van Tend} and {Kuperus}(1978)}]{vanTend1978a}
{Van Tend}, W. and {Kuperus}, M. (1978).
\newblock The development of coronal electric current systems in active regions
  and their relation to filaments and flares.
\newblock \emph{\solphys} 59, 115--127.
\newblock \doi{10.1007/BF00154935}
\input{vanTend1978a}

\bibitem[{{Vr{\v s}nak} et~al.(2005){Vr{\v s}nak}, {Sudar}, and {Ru{\v
  z}djak}}]{Vrvsnak2005a}
{Vr{\v s}nak}, B., {Sudar}, D., and {Ru{\v z}djak}, D. (2005).
\newblock The cme-flare relationship: Are there really two types of cmes?
\newblock \emph{\aap} 435, 1149--1157.
\newblock \doi{10.1051/0004-6361:20042166}
\input{Vrvsnak2005a}

\bibitem[{Wang et~al.(2007)Wang, Sui, and Qiu}]{Wang2007}
Wang, T., Sui, L., and Qiu, J. (2007).
\newblock Direct observation of high-speed plasma outflows produced by magnetic
  reconnection in solar impulsive events.
\newblock \emph{\apjl} 661, L207--L210.
\newblock \doi{10.1086/519004}
\input{Wang2007}

\bibitem[{Xia et~al.(2012)Xia, Chen, and Keppens}]{Xia2012}
Xia, C., Chen, P.~F., and Keppens, R. (2012).
\newblock Simulations of prominence formation in the magnetized solar corona by
  chromospheric heating.
\newblock \emph{\apjl} 748, L26.
\newblock \doi{10.1088/2041-8205/748/2/L26}
\input{Xia2012}

\bibitem[{Xue et~al.(2016)Xue, Yan, Cheng, Yang, Su, Kliem et~al.}]{Xue2016}
Xue, Z., Yan, X., Cheng, X., Yang, L., Su, Y., Kliem, B., et~al. (2016).
\newblock Observing the release of twist by magnetic reconnection in a solar
  filament eruption.
\newblock \emph{Nature Communications} 7, 11837.
\newblock \doi{10.1038/ncomms11837}
\input{Xue2016}

\bibitem[{Yan et~al.(2022)Yan, Xue, Jiang, Priest, Kliem, Yang
  et~al.}]{Yan2022}
Yan, X., Xue, Z., Jiang, C., Priest, E.~R., Kliem, B., Yang, L., et~al. (2022).
\newblock Fast plasmoid-mediated reconnection in a solar flare.
\newblock \emph{Nature Communications} 13, 640.
\newblock \doi{10.1038/s41467-022-28269-w}
\input{Yan2022}

\bibitem[{{Zhang} et~al.(2001){Zhang}, {Dere}, {Howard}, {Kundu}, and
  {White}}]{Zhang2001}
{Zhang}, J., {Dere}, K.~P., {Howard}, R.~A., {Kundu}, M.~R., and {White}, S.~M.
  (2001).
\newblock {On the Temporal Relationship between Coronal Mass Ejections and
  Flares}.
\newblock \emph{\apj} 559, 452--462.
\newblock \doi{10.1086/322405}
\input{Zhang2001}

\bibitem[{Zhang et~al.(2021)Zhang, Liu, Wang, Li, and Lyu}]{Zhang2021}
Zhang, Q., Liu, R., Wang, Y., Li, X., and Lyu, S. (2021).
\newblock Confined and eruptive catastrophes of solar magnetic flux ropes
  caused by mass loading and unloading.
\newblock \emph{\apj} 921, 172.
\newblock \doi{10.3847/1538-4357/ac1fef}
\input{Zhang2021}

\bibitem[{{Zhang} et~al.(2017{\natexlab{a}}){Zhang}, {Wang}, {Hu}, {Liu}, and
  {Liu}}]{Zhang2017a}
{Zhang}, Q., {Wang}, Y., {Hu}, Y., {Liu}, R., and {Liu}, J.
  (2017{\natexlab{a}}).
\newblock Influence of photospheric magnetic conditions on the catastrophic
  behaviors of flux ropes in solar active regions.
\newblock \emph{\apj} 835, 211.
\newblock \doi{10.3847/1538-4357/835/2/211}
\input{Zhang2017a}

\bibitem[{{Zhang} et~al.(2017{\natexlab{b}}){Zhang}, {Wang}, {Hu}, {Liu},
  {Liu}, and {Liu}}]{Zhang2017}
{Zhang}, Q., {Wang}, Y., {Hu}, Y., {Liu}, R., {Liu}, K., and {Liu}, J.
  (2017{\natexlab{b}}).
\newblock {Upward and Downward Catastrophes of Coronal Magnetic Flux Ropes in
  Quadrupolar Magnetic Fields}.
\newblock \emph{\apj} 851, 96.
\newblock \doi{10.3847/1538-4357/aa9ce6}
\input{Zhang2017}

\bibitem[{Zhang et~al.(2020)Zhang, Wang, Liu, Zhang, Hu, Wang
  et~al.}]{Zhang2020}
Zhang, Q., Wang, Y., Liu, R., Zhang, J., Hu, Y., Wang, W., et~al. (2020).
\newblock Eruption of solar magnetic flux ropes caused by flux feeding.
\newblock \emph{\apjl} 898, L12.
\newblock \doi{10.3847/2041-8213/aba1f3}
\input{Zhang2020}

\end{thebibliography}

\end{document}